\documentstyle[A4,11pt,epsfig,twoside]{article}
\def\etal{{\it et al.}}
\begin{document}
\setcounter{page}{1}
\title{Charged Higgs and Stau Production in Anomaly Mediated 
Supersymmetry Breaking
}
\author{M.A. D\'IAZ\thanks{Speaker.
Presented at Workshop on Physics and Experiments with Future 
Electron-Positron Linear Colliders August 26-30, 2002, Jeju Island, Korea.
}\,\,, R.A. LINEROS, AND M.A. RIVERA
\\
\\
{\it Departamento de F\'\i sica, Universidad Cat\'olica de Chile,}
\\
{\it Av. Vicu\~na Mackenna 4860, Santiago, Chile}
}
\date{}
\maketitle
\begin{abstract}
Charged Higgs production in association with staus in electron positron 
annihilation is a signal of supersymmetry with bilinear R-Parity violation.
In this model, neutrino masses and mixing angles are generated due to 
mixing with neutralinos. We show how parameters related to neutrino 
physics can be determined at a $500\,{\mathrm fb}^{-1}$ Linear Collider 
from measurements of charged Higgs and stau production cross sections 
and decay rates. This can be achieved in AMSB where charged Higgs and 
stau can have similar masses.
\end{abstract}


In supersymmetric models which include Bilinear R-Parity Violation 
(BRpV)\cite{BRpV}, the charged Higgs boson field mixes with the charged 
sleptons and as a result their phenomenology changes 
dramatically\cite{Akeroydetal}. In particular, charged Higgs and staus
can be produced in association in electron positron collisions.
This mixed production is a signal of BRpV since it is not present in 
models with only trilinear R-Parity violating couplings\cite{Rpreview}.
The associated production cross section is usually small, nevertheless,
a near degeneracy of the scalars can enhance it. This is possible in
Anomaly Mediated Supersymmetry Breaking (AMSB) models\cite{AMSB}, as it was
shown in\cite{Brazil}. Here we show that using measurements of 
production cross sections, decay rates, and masses in the charged scalar 
sector taken at a future Linear Collider\cite{LC}, it is possible to 
extract values for the parameters associated to neutrino physics\cite{DLR}.

In BRpV-MSSM three parameters that violate R-Parity and lepton number 
are included in the superpotential,
\begin{equation}
W=W_{MSSM}+\epsilon_i\widehat L_i\widehat H_u
\end{equation}
The $\epsilon_i$ parameters have units of mass and they induce sneutrino 
vacuum expectation values $v_i$. As a result, neutrino masses and mixing 
angles are generated by a low energy see-saw mechanism due to mixing with 
neutralinos. At tree level the atmospheric mass scale is given by
\begin{equation}
\label{mnutree}
\sqrt{\Delta m_{atm}^2}\approx m_{\nu_3}= 
\frac{M_1 g^2 + M_2 g'^2}{4\, det({\cal M}_{\chi^0})} 
|{\vec \Lambda}|^2.
\end{equation}
where, $\Lambda_i=\mu v_i+\epsilon_i v_d$. It has been shown that 
information on these parameters can be extracted from LSP decays, either a
neutralino\cite{Romaoetal}, or a stau\cite{LSPstau}. This is possible 
despite the fact the BRpV are small, because the LSP decays 100\% into SM
particles. Here we show that in AMSB, R-Parity violating observables in 
the charged scalar sector not necessarily related to LSP decays can also
be measurable. This is because charged Higgs and stau mixings, which are
of the form\cite{DLR},
\begin{equation}
s_i={{X_{H\tilde\tau_i}}\over{m_{H^{\pm}}^2-m_{\tilde\tau_i}^2}}
\end{equation}
where $X_{H\tilde\tau_i}$ is the mixing in the mass matrix, can be 
enhanced when there is near degeneracy.

In this context we perform a $\chi^2$ analysis whose input is an AMSB-BRpV
benchmark scenario given by $M_{3/2}=30$ TeV, $m_0=200$ GeV, $\tan\beta=15$,
$\mu<0$, $\epsilon_3=1$ GeV, and $m_{\nu}=0.1$ eV. In this scenario the 
charged Higgs with $m_{H^{\pm}}=188$ GeV is nearly degenerate with the heavy
stau with $m_{\tilde\tau_2}=190$ GeV (mass differences of the order of 
10 GeV still would be useful for our purposes). The associated production 
cross sections at $\sqrt{s}=1$ TeV are reasonably large with 
$\sigma(e^+e^-\rightarrow H^{\pm}\tilde\tau_1^{\mp})=1.81$ fb and
$\sigma(e^+e^-\rightarrow H^{\pm}\tilde\tau_2^{\mp})=0.79$, producing a few
hundred events with an integrated luminosity of 
${\cal L}=500\,{\mathrm fb}^{-1}$. In this analysis we include only the 
statistical error for the cross sections and an estimated error of 1\% for 
the masses.

\begin{figure}
\centerline{\protect\hbox{\epsfig{file=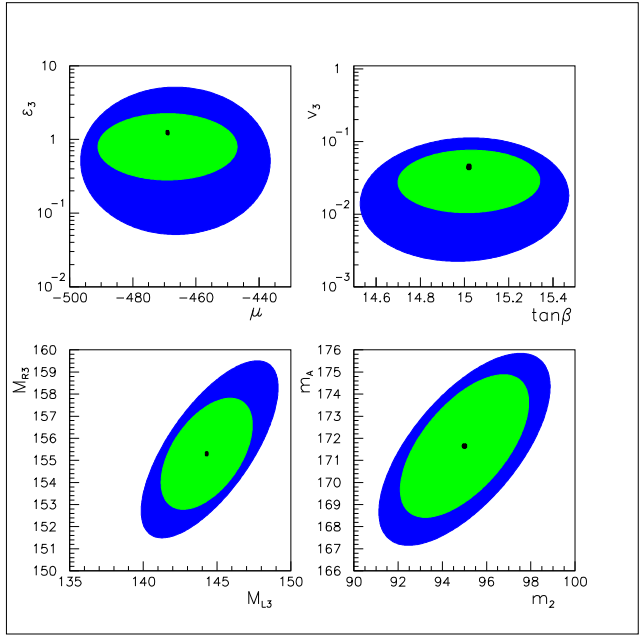,width=0.6\textwidth}}}
\vskip .5cm
\caption[]{
Regions of parameter space where normalized $\chi^2\le 1$ are shown in 
green (light gray). For comparison, also shown are the 
$\chi^2\le 2$ regions in blue (dark gray).
}
\label{chi2_sca}
\end{figure}
The output of the $\chi^2$ analysis can be seen in Fig.~\ref{chi2_sca},
where we plot regions with $\chi^2<1$. The width of these
ellipses indicate the error on the determination of each parameter.
Our findings are summarized in Table~\ref{tab:cases}, from which we 
highlight $\epsilon_3=1.24\pm0.96$ GeV and $v_3=45\pm35$ MeV. Despite the 
error, this determination is very encouraging because it is hard to 
extract them from other observables.

\begin{table}
\begin{center}
\caption{Parameter determination from charged Higgs and stau associated 
production and decay. All parameters 
are expressed in GeV, except for $\tan\beta$.
}
\bigskip
\begin{tabular}{lccccc}
\hline
parameter & input & output  & error & percent  \\
\hline

$\epsilon_3 $ & 1.0   & 1.24   & 0.96    & 77     \\
$v_3        $ & 0.035 & 0.045  & 0.035   & 78     \\
$\tan\beta  $ & 15.0  & 15.0   & 0.3     &  2     \\
$\mu        $ & -466  & -469   & 22      &  5     \\
$M_{L_3}    $ & 155.8 & 155    & 3       &  2     \\
$M_{R_3}    $ & 144.5 & 144    & 3       &  2     \\
$m_A        $ & 171.5 & 172    & 3       &  2     \\
$M_2        $ & 95.4  & 95     & 3       &  3     \\
\hline
\label{tab:cases}
\end{tabular}
\end{center}
\end{table}

In summary, we have shown that it is possible to extract, although with 
a large error, the value of the BRpV parameters $\epsilon_3$ present in 
the superpotential and the sneutrino vacuum expectation value $v_3$
from charged Higgs and stau associated production at the LC, where 
precision measurements can be performed. This is possible in AMSB 
model in the region of parameter space where 
the charged Higgs and the stau have similar mass, which enhances the 
mixing between these two fields. This has important consequences in 
neutrino physics, since these parameters define the atmospheric and solar 
mass scales and mixing angles.

\end{document}